\documentstyle[12pt]{article}
\catcode`\@=11

\@addtoreset{equation}{section}
\def\theequation{\arabic{equation}}

\def\@normalsize{\@setsize\normalsize{15pt}\xiipt\@xiipt
\abovedisplayskip 14pt plus3pt minus3pt%
\belowdisplayskip \abovedisplayskip
\abovedisplayshortskip  \z@ plus3pt%
\belowdisplayshortskip  7pt plus3.5pt minus0pt}
\def\small{\@setsize\small{13.6pt}\xipt\@xipt
\abovedisplayskip 13pt plus3pt minus3pt%
\belowdisplayskip \abovedisplayskip
\abovedisplayshortskip  \z@ plus3pt%
\belowdisplayshortskip  7pt plus3.5pt minus0pt
\def\@listi{\parsep 4.5pt plus 2pt minus 1pt
            \itemsep \parsep
            \topsep 9pt plus 3pt minus 3pt}}

\def\underline#1{\relax\ifmmode\@@underline#1\else
        $\@@underline{\hbox{#1}}$\relax\fi}
\@twosidetrue
\relax

\catcode`@=12

\catcode`\@=11

\def\section{\@startsection{section}{1}{\z@}{3.5ex plus 1ex minus
   .2ex}{2.3ex plus .2ex}{\large\bf}}

\def\ps@headings{\def\@oddfoot{}\def\@evenfoot{}
\def\@oddhead{\hbox{}\hfill
        \makebox[.5\textwidth]{\raggedright\ignorespaces --\thepage{}--
        \hfill }}
\def\@evenhead{\@oddhead}
\def\subsectionmark##1{\markboth{##1}{}}
}

\ps@headings

\catcode`\@=12

\relax

\def\figcap{\section*{Figure Captions\markboth
        {FIGURECAPTIONS}{FIGURECAPTIONS}}\list
        {Fig. \arabic{enumi}:\hfill}{\settowidth\labelwidth{Fig. 999:}
        \leftmargin\labelwidth
        \advance\leftmargin\labelsep\usecounter{enumi}}}
 \relax
\def\tablecap{\section*{Table Captions\markboth
        {TABLECAPTIONS}{TABLECAPTIONS}}\list
        {Table \arabic{enumi}:\hfill}{\settowidth\labelwidth{Table 999:}
        \leftmargin\labelwidth
        \advance\leftmargin\labelsep\usecounter{enumi}}}
 \relax
\def\reflist{\section*{References\markboth
        {REFLIST}{REFLIST}}\list
        {[\arabic{enumi}]\hfill}{\settowidth\labelwidth{[999]}
        \leftmargin\labelwidth
        \advance\leftmargin\labelsep\usecounter{enumi}}}
 \relax

\catcode`\@=11

\def\marginnote#1{}
\newcount\hour
\newcount\minute
\newtoks\amorpm
\hour=\time\divide\hour by60
\minute=\time{\multiply\hour by60 \global\advance\minute by-
\hour}
\edef\standardtime{{\ifnum\hour<12 \global\amorpm={am}%
    \else\global\amorpm={pm}\advance\hour by-12 \fi
    \ifnum\hour=0 \hour=12 \fi
    \number\hour:\ifnum\minute<100\fi\number\minute\the\amorpm}}
\edef\militarytime{\number\hour:\ifnum\minute<100\fi\number\minute}
\def\draftlabel#1{{\@bsphack\if@filesw {\let\thepage\relax
  \xdef\@gtempa{\write\@auxout{\string
    \newlabel{#1}{{\@currentlabel}{\thepage}}}}}\@gtempa
    \if@nobreak \ifvmode\nobreak\fi\fi\fi\@esphack}
     \gdef\@eqnlabel{#1}}
\def\@eqnlabel{}
\def\@vacuum{}
\def\draftmarginnote#1{\marginpar{\raggedright\scriptsize\tt#1}}
\def\draft{\oddsidemargin -.5truein
        \def\@oddfoot{\sl preliminary draft \hfil
        \rm\thepage\hfil\sl\today\quad\militarytime}
        \let\@evenfoot\@oddfoot \overfullrule 3pt
        \let\label=\draftlabel
        \let\marginnote=\draftmarginnote
   
\def\@eqnnum{(\theequation)\rlap{\kern\marginparsep\tt\@eqnlabel}%
\global\let\@eqnlabel\@vacuum}  }
\def\preprint{\twocolumn\sloppy\flushbottom\parindent 1em
        \leftmargini 2em\leftmarginv .5em\leftmarginvi .5em
        \oddsidemargin -.5in    \evensidemargin -.5in
        \columnsep 15mm \footheight 0pt
        \textwidth 250mmin      \topmargin  -.4in
        \headheight 12pt \topskip .4in
        \textheight 175mm
        \footskip 0pt
        
\def\@oddhead{\thepage\hfil\addtocounter{page}{1}\thepage}
        \let\@evenhead\@oddhead \def\@oddfoot{} \def\@evenfoot{} 
}
\def\titlepage{\@restonecolfalse\if@twocolumn\@restonecoltrue\onecolumn
     \else \newpage \fi \thispagestyle{empty}\c@page\z@
        \def\thefootnote{\fnsymbol{footnote}} }
\def\endtitlepage{\if@restonecol\twocolumn \else  \fi
        \def\thefootnote{\arabic{footnote}}
        \setcounter{footnote}{0}}  %\c@footnote\z@ }
\catcode`@=12
\relax
\def\ps@headings{\def\@oddfoot{}\def\@evenfoot{}
\def\@oddhead{\hbox{}\hfill
        \makebox[.5\textwidth]{\raggedright\ignorespaces --\thepage{}--
        \hfill }}
\def\@evenhead{\@oddhead}
\def\subsectionmark##1{\markboth{##1}{}}
}

\ps@headings

\relax

\def\firstpage#1#2#3#4#5#6{
\begin{document}
%\draft
%%%%%%%%%%%%%%%%% MACROS %%%%%%%%%%%%%%%%%%
\def\beq{\begin{equation}} 
\def\eeq{\end{equation}} 
\def\bea{\begin{eqnarray}} 
\def\eea{\end{eqnarray}} 
\def\bq{\begin{quote}} 
\def\eq{\end{quote}}
\def\ra{\rightarrow} 
\def\lra{\leftrightarrow} 
\def\ups{\upsilon}
\def\bq{\begin{quote}} 
\def\eq{\end{quote}}
\def\ra{\rightarrow} 
\def\un{\underline}
\def\ov{\overline}
\newcommand{\cm}{Commun.\ Math.\ Phys.~}
\newcommand{\prl}{Phys.\ Rev.\ Lett.~}
\newcommand{\pr}{Phys.\ Rev.\ D~}
\newcommand{\pl}{Phys.\ Lett.\ B~}
\newcommand{\ibar}{\bar{\imath}}
\newcommand{\jbar}{\bar{\jmath}}
\newcommand{\np}{Nucl.\ Phys.\ B~}
\newcommand{\F}{{\cal F}}
\renewcommand{\L}{{\cal L}}
\newcommand{\A}{{\cal A}}
\def\154{\frac{15}{4}}
\def\153{\frac{15}{3}}
\def\32{\frac{3}{2}}
\def\254{\frac{25}{4}}
\begin{titlepage}
\nopagebreak
\title{\begin{flushright}
        \vspace*{-1.8in}
        %{\normalsize CERN-TH/98-nnn}\\[-9mm]
        {\normalsize DEMO-HEP.98-01}\\[-9mm]
        {\normalsize IOA--TH.98-01}\\[-9mm]
        %{\normalsize UUTP-22/97}\\[-9mm]
        {\normalsize hep-th/9802018}\\[4mm]
\end{flushright}
\vfill
{#3}}
\author{\large #4 \\[1.0cm] #5}
\maketitle
\vskip -7mm     
\nopagebreak 
\begin{abstract}
{\noindent #6}
\end{abstract}
\vfill
\begin{flushleft}
\rule{16.1cm}{0.2mm}\\[-3mm]
%$^{\star}${\small %thanks etc }\\ 
%$^{\dagger}${\small Supported by the European Community under Human 
%Capital and Mobility Grant No ERBCHBICT960773}\\
%CERN-TH/98-nnn\\
February 1998
\end{flushleft}
\thispagestyle{empty}
\end{titlepage}}
\def\simlt{\stackrel{<}{{}_\sim}}
\def\simgt{\stackrel{>}{{}_\sim}}
\date{}
\firstpage{3118}{IC/95/34}
{\large\bf A Note on the Supersymmetries of the Self-Dual Supermembrane} 
{E.G. Floratos$^{\,a,b}$ and  G.K. Leontaris$^{\,c}$
}%\\[-3mm] 
{\normalsize\sl
$^a$ NRCS Demokritos, Athens, Greece\\[-3mm]
\normalsize\sl
$^b$ Physics Department, University of Crete, Iraklion,
Crete, Greece.\\[-3mm]
\normalsize\sl
$^c$Theoretical Physics Division, Ioannina University,
GR-45110 Ioannina, Greece.}
{In this letter we discuss the supersymmetry issue  of the self-dual
supermembranes in $(8+1)$ and $(4+1)$-dimensions. We find that all 
genuine solutions of the $(8+1)$-dimensional supermembrane, based on the 
exceptional group $G_2$, preserve one of the sixteen supersymmetries
while all solutions in $(4+1)$-dimensions preserve eight of them. 
}

%%%%%%%%%%%%%%%

\newpage

Recently,  a new  duality for fundamental membranes~\cite{FL0} 
in ($4+1$)-dimensions, has been extended to $(8+1)$-dimensions
using the structure constants of the octonionic algebra~\cite{1,2,3}.
Explicit solutions have been constructed in various dimensions and
connections with string instantons have been found~\cite{4}. 

{}The fundamental supermembranes as extended objects were first 
described in~\cite{BST} by a manifestly space-time supersymmetric
Green-Schwarz (GS)-action. It was further shown~\cite{DS} that they 
emerge as a solution of the eleven dimensional  supergravity field 
equations with their zero modes corresponding to the physical degrees 
of freedom of the GS-action.
It is now known that one of the fundamental problems of supermembrane
theory is the existence of a convenient perturbative expansion and 
the derivation of effective low energy Lagrangian (which is 
expected to be the 11-dimensional $N=1$ supergravity theory).
{}For this problem, the existence of a self-dual sector of 
BPS-states for the supermembrane, preserving a number of
supersymmetries which would guarantee the absence of
perturbative corrections could be a way out. An interesting 
property of the self-duality equations for supermembrane is that 
in three dimensions the system is an integrable one and in principle 
all the spectrum of the corresponding BPS-states could be determined. 
In this case, after the light-cone gauge fixing, one restricts the 
membrane to the three of the nine dimensions in order to formulate 
the self-duality equations. The corresponding integrability
of the seven-dimensional case is still under investigation.

In this letter, we study the supersymmetry transformations for the 
octonionic self-dual membranes and we determine the number of 
supersymmetries left in seven and three dimensions.  We find 
that the seven-dimensional case preserves one supersymmetry,
while the three-dimensional solutions preserve eight of them.
The $G_2$ symmetry of the seven-dimensional case can be used 
to embed the $N=8$, $d=3$ BPS-states into $N=8$, $d=7$ superalgebra.

We start by recalling the light-cone gauge formulation of the 
supermembrane,  where half of the rigid space-time supersymmetry 
as well as the local $\kappa$-symmetry is fixed. We then provide  
the supersymmetries left intact. 

It is known that in $d=8$ dimensions there is a connection of the 
Clifford algebra with the octonionic algebra and this is the information 
needed to study the behaviour of the octonionic self-duality equations 
under supersymmetry transformations. The relation with the octonions has 
been noticed in the 80's during the studies of the $S^7$-compactifications 
of the 11-d supergravity as well as for the $N=8$ gauged 
supergravities~\cite{5aa,5a,ivan}. Recently, the embedding of octonionic
Yang-Mills (YM) instantons in the ten-dimensional effective supergravity
theories of strings has been constructed~\cite{6,7,7a} where it was found
that one supersymmetry survives.  More generally, wrapped membrane
compactifications have been recently discussed in the literature~\cite{wrap}.

In the light-cone gauge, after the elimination of the $X_-$ variable from the
reparametrization constraints, the supersymmetric Hamiltonian~\cite{BST,NdW}
is defined as
\bea
{\cal H} = \frac 1{P_0^+}\int d^2\sigma\left(
                  \frac 12 P^IP_I+\frac 14 \{X^I,X^J\}^2
           -P_0^+\bar{\theta}\Gamma_-\Gamma_I\{X^I,\theta\}\right)
\label{Ham}
\eea
where $P_I=\dot{X}_I$ and  the indices $I,J =1,...,9$ while we have fixed 
the area preserving parameters so that $w=1$~\cite{BST}. The compatibility  
condition for the uniqueness of $X_-$, is the  Gauss law
\bea
\{\dot{X}_I,X_I\}+\{\bar\theta\Gamma_-,\theta\}=0,\,\, I=1,...,9.
\label{GL}
\eea
where summation over repeated incides is assumed. The Clifford 
generators $\Gamma_I$, in (\ref{Ham}) are represented by real 
$32\times 32$ matrices which can be chosen in the following form 
\bea
& \Gamma_{I} = \sigma_3\otimes \gamma_{I} 
\label{Gamma}
\eea
where $\sigma_3$ is the Pauli matrix, $\gamma_{I}$  represent  the
$16\times 16$ matrices and $\gamma_9=\gamma_1\cdots \gamma_8$. Further,
$\Gamma_-$ (and $\Gamma_+$) correspond to the light-cone coordinates
($X_{10}\pm X_0)/\sqrt{2}$, thus they are given by a similar decomposition 
\bea
\Gamma_{\pm}=\frac 1{\sqrt{2}} \left(\Gamma_{10}\pm\Gamma_0\right).
\label{Gpm}
\eea
Thus, we have
\bea
\Gamma_{-}=\imath\sqrt{2}
\left(\begin{array}{cc}
0 & 1_{16} \\
0&0
\end{array} \right),&
\Gamma_{+}=\imath\sqrt{2}
\left(\begin{array}{cc}
0 & 0 \\
1_{16}&0
\end{array} \right)
\label{Gpm1}
\eea

The Hamiltonian (\ref{Ham}) is invariant under area-preserving transformations 
of the membrane (which for non-trivial topologies of the membrane contain also 
global elements $2 g$ in number, where $g$ is the genus of the membrane~\cite{NdW}). 
The local area-preserving transformations are generated by the Gauss law (\ref{GL}).
Here the canonical variables satisfy Dirac brackets
\bea
\left(X^I(\sigma),\dot{X}^J(\sigma')\right)_{DB}=\delta^{IJ}\delta^2
(\sigma -\sigma')\\
\left(\theta^I(\sigma),\bar\theta^J(\sigma')\right)_{DB}=
\frac 14{(\Gamma_+)}^{IJ}\delta^2(\sigma -\sigma')
\eea
(where we have chosen $P_+=1$).
It can be verified that in the light-cone gauge there are two independent
spinor supersymmetry charges
\bea
Q=Q^++Q^-=\int d^2\sigma J^0
\eea
where $Q^{\pm}=\frac 12\Gamma_{\pm}\Gamma_{\mp}Q$, and 
\bea
Q^+&=& \int d^2\sigma (2\dot{X}^I\Gamma_I+\{X^I,X^J\}\Gamma_{IJ})
\theta\\
Q^-&=& \-2\int d^2\sigma S = 2 \Gamma_-\theta_0
\eea
which are constants of motion and $\theta_0$ is the momentum conjugate to the 
center-of-mass coordinate of the fermionic degrees of freedom of the membrane.
The corresponding supersymmetry transformations which leave the Hamiltonian
invariant, are given by
\bea
\delta X^I& =& -2\bar{\epsilon}\Gamma^I\theta\\
\delta\theta &=& \frac 12\Gamma_+\left(\dot{X}\Gamma_I+\Gamma_-\right)
\epsilon +\frac 14\{X^I,X^J\}\Gamma_+\Gamma^{IJ}\epsilon
\eea
On the other hand, the local fermionic $\kappa$-symmetry has been fixed
by imposing the condition
\beq
\Gamma_+\theta = 0
\eeq
Due to this gauge condition, the fermionic coordinates are restricted to
$SO(9)$ spinors, satisfying 
\beq
\Gamma_1\cdots \Gamma_9 \theta=\theta
\eeq
while the $SO(9)$ $\Gamma$-matrices satisfy $\Gamma_I^T=\Gamma^I$.

The self-duality equations for the bosonic
part of the supermembrane have been initially introduced in the 
light-cone gauge fixing $X_4,\dots X_9$  to be constants~\cite{FL0},
\beq
\dot{X}_i = \frac 12\epsilon_{ijk}\{X_j,X_k\},
\;\; i,j=1,2,3
\label{sde}
\eeq
These equations have been proposed as an analogue of the
electric-magnetic duality where the local velocity of the membrane
corresponds to the electric field while the RHS which is the normal
to the membrane surface, corresponds to the magnetic field.
They imply the Gauss law and the Euclidean-time  equations 
of motion with fermionic degrees of freedom (dof) set to zero~\cite{FL0,W}.
This system has been shown to be integrable and a Lax pair was found.
In order to go to higher dimensions one should have the notion of
cross product of two vectors and this is provided as the unique 
other possibility by the structure constants of the algebra of 
octonions (Cayley numbers)~\cite{Gursey}.
The octonionic units $o_i$ satisfy the algebra
\beq
o_i o_j = -\delta_{ij} + \Psi_{ijk} o_k.
\label{omult}
\eeq
where $i=1,\dots , 7$ are the 7 octonionic imaginary units with the property
\beq
\{o_i,o_j\} = - 2\delta_{ij}
\eeq
{}The totally antisymmetric symbol $\Psi_{ijk}$ appearing in (\ref{omult})
is defined to be equal to 1 when the indices are~\cite{Gursey}
\beq
\Psi_{ijk}=\left\{\begin{array}{ccccccc}1&2&4&3&6&5&7\\
                             2&4&3&6&5&7&1\\
                             3&6&5&7&1&2&4
\end{array}\right.
\label{2.1}
\eeq
and zero for all other cases. With this multiplication table, $\Psi_{ijk}$ 
provides for every two seven-dimensional vectors a third one, normal to the 
first two. Thus, it is possible to extend the three-dimesional self-duality 
equations to seven dimensions, fixing only the values of $X_8,X_9$ membrane 
coordinates. Then, the self-duality equations~\cite{1}  become
\begin{equation}
\dot{X}_i = \frac{1}{2} \Psi_{ijk}\{X_j,X_k\}
\label{osce}
\end{equation}
The Gauss law results automatically by making use of the $\Psi_{ijk}$ cyclic
symmetry
\beq
\{\dot{X}_i,X_i\}= 0
\eeq
The Euclidean equations of motion  are obtained easily from (\ref{osce})
\bea
\ddot{X}_i& =& 
%\frac 12\Psi_{ijk}\left(\{\dot{X}_j,X_k\}+ 
%\{X_j,\dot{X}_k\}\right)\\
%          & =  &
          \{X_k,\{X_i,X_k\}\}
\eea
where use has been made of the identity
\beq
\Psi_{ijk}\Psi_{lmk}=\delta_{il}\delta_{jm}-\delta_{im}\delta_{jl}+\phi_{ijlm}
\label{I1}
\eeq
and  of the cyclic property of the symbol
$\phi_{ijlm}$~\cite{Gursey} which 
is defined to be equal to 1 when its indices take   
values of the following table 
\beq
{\phi^{ij}}_{kl}=\left\{\begin{array}{ccccccc}
                             4&3&6&5&7&1&2\\
                             5&7&1&2&4&3&6\\
                             6&5&7&1&2&4&3\\
                             7&1&2&4&3&6&5
\end{array}\right. 
\label{phi}
\eeq
whilst it is zero for any other combination of indices.
In terms of these units an octonion  can be written as follows
\beq
{X} = x_0 o_0 + \sum_{i=1}^7 x_io_i
\eeq
with $o_0$ the identity element. The  conjugate is
\beq
\bar{X} = x_0 o_0 -  \sum_{i=1}^7 x_io_i
\eeq
The  octonions over the real numbers 
 can also be defined as pairs of quaternions
\beq
X = (x_1,x_2)\label{o-q}
\eeq
where $x_{1}= x_1^{\mu}\sigma_{\mu}$, $x_2 =x_2^{\mu}\sigma_{\mu}$ and the indices 
$\mu $ run from $0$ to 3, while $x_{1,2}^0$ are real numbers and  $x_{1,2}^i,
\; (i=1,2,3)$ are imaginary numbers. Finally, $\sigma_{0}$ is the identity 
$2\times 2$ matrix and  $\sigma_{i}$ are the three standard Pauli matrices.

If $q = (q_1,q_2)$ and $r=(r_1,r_2)$ are two octonions, the multiplication law is 
defined as
\beq
q*r \equiv (q_1,q_2)*(r_1,r_2) = (q_1r_1-\bar{r}_2q_2, r_2q_1+q_2\bar{r}_1),
\label{mr}
\eeq
where $q_1 = q_1^0+q_1^i\sigma_i$ and $\bar{q}_1= q_1^0-q_1^i\sigma_i$.
One can also define a conjugate operation for an octonion as
\beq
\bar{q} \equiv {\overline{(q_1,q_2)}} =(\bar{q}_1,-q_2)
\eeq
and we get the possibility to define  the norm and the scalar product
$q$~and~$r$
\bea
q\bar{q} & =& (q_1\bar{q}_1+\bar{q}_2q_2,0)\nonumber\\
         &= & \sum_{\mu=0}^3\left({q_1^{\mu}}^2+{q_2^{\mu}}^2\right)\\
\langle q|r\rangle &=& \frac 12 (q\bar{r}+\bar{q}r)
\eea
In terms of the above formalism, the self-duality equations can be written 
as follows
\beq
\dot{X}= \frac 12\{X,X\}\label{sdo},
\eeq
where $X=X^io_i$ with $i=1,\cdots , 7$ and the Poisson bracket 
for two octonions is defined as
\beq
\{X,Y\} =\frac{\partial{X}}{\partial{\xi_1}}\frac{\partial{Y}}{\partial{\xi_2}}
   -\frac{\partial{X}}{\partial{\xi_2}}\frac{\partial{Y}}{\partial{\xi_1}}.
\eeq

After these preliminaries, we come now to the question regarding the
number of supersymmetries preserved by the self-duality equations.
In our analysis we will explore the number of supersymmetries preserved
by (3+1)- and (7+1)-dimensional solutions. We will see that 3-d solutions
preserve as many as eight out of the sixteen supersymmetries while the
7-d self-duality equations preserve only one supersymmetry. The supersymmetry 
transformation is defined~\cite{BST}
\bea
\delta\theta &=&\frac 12\left( \Gamma_+(\Gamma_I\dot{X}^I+\Gamma_-)+
                  \frac 12\Gamma_+\Gamma^{IJ}\{X_I,X_J\}\right)
\left(\begin{array}{c}
\imath\epsilon_A\\
\epsilon_B\end{array} \right)\nonumber
\eea
In terms of the $16\times 16$ $\gamma$-matrices, the above is written
\bea
\delta\theta &=& \left(\begin{array}{cc}
0 & 0\\
\imath\sqrt{2}\left(\gamma^I\dot{X}_I+\frac 12\gamma^{IJ}
\{X_I,X_J\}\right)&-2\cdot 1_{16}
\end{array} \right)\left(\begin{array}{c}
\imath\epsilon_A\\
\epsilon_B\end{array} \right)\label{delthe32}
\eea
which implies that 
\bea
\sqrt{2}\left(\gamma^I\dot{X}_I+
\frac 12\gamma^{IJ}\{X_I,X_J\}\right)\epsilon_A
+ 2\cdot 1_{16} \epsilon_B=0\label{delthe16}
\eea
where $ \epsilon_A,  \epsilon_B$ are 16-dimensional spinors. {}From  
the form of eq.(\ref{delthe16}), we  observe that if self-duality 
equations are going to play a role in the preservation of a number of 
supersymmetries, we should necessarily impose the condition $\epsilon_B=0$.
Thus, at least half of the supersymmetries are broken.  Now, the last term 
in (\ref{delthe16}) is zero and eq.(\ref{delthe16}) simply becomes
\bea
\left(\gamma^I\dot{X}_I+
\frac 12\gamma^{IJ}\{X_I,X_J\}\right)\epsilon_A = 0\label{16SD}
\label{delthe16A}
\eea
Under the assumption that $\dot{X}_{8,9}=0$, it can be  shown that the above  
reduces to a simpler --$8\times 8$-- matrix equation. In order to find a 
convenient explicit form, we first express the $16\times 16$ matrices in 
terms of the octonionic structure constants $\Psi_{ijk}$ as follows: 
let the index $n$ run from 1 to 7; then we define
\bea
\gamma_{8} = 
\left(\begin{array}{cc}
0 & 1_8 \\
-1_8&0 
\end{array} \right),&
\gamma_{n} = 
\left(\begin{array}{cc}
0 & \beta_n \\
-\beta_n&0 
\end{array} \right) \label{gn}
\eea
where $1_8$ is the $8\times 8$-identity matrix and $\beta_n$ are 
seven  $8\times 8$  $\gamma$-matrices with elements~\cite{ivan}
\bea
(\beta_n)^i_j = \Psi_{imj},& (\beta_n)^i_8=\delta^i_j,
&(\beta_n)^8_j=-\delta^i_j
\label{belem}
\eea
while it can be easily checked that $\beta_1\cdots\beta_7=-1_8$ and
\bea
\gamma_9 % &=&\gamma_1\cdots\gamma_8\nonumber\\
&=& \left(\begin{array}{cc}
 1_{8}&0 \\
0&-1_8
\end{array} \right)
\eea
The commutation relations of $\beta_m$ give:
\bea
\left([\beta_m,\beta_n]\right)^8_j&=& +2 \Psi_{nmj}
\\
\left([\beta_m,\beta_n]\right)^j_8&=&-2 \Psi_{nmj}
\\
\left([\beta_m,\beta_n]\right)^i_j&=&-2{{\cal X}^{mn}}_{ij}(-4)
\eea
where the tensors ${{\cal X}^{mn}}_{ij}(u)$ are defined as follows~\cite{2}
\beq
{{\cal X}^{ij}}_{kl}(u) = {\Delta^{ij}}_{kl}+\frac u4{\phi^{ij}}_{kl}
\eeq
where ${\Delta^{ij}}_{kl}=\frac 12(\delta_k^i\delta_l^j-\delta_l^i\delta_k^j)$.
Next, we impose the following condition on the components of the 16-spinor
$\epsilon_A$ 
\bea
\epsilon_A&=& \left(\begin{array}{c}1\\
                                    -\imath
              \end{array} \right)   
 \otimes            
             \varepsilon\label{8spin}
\eea
where $\otimes$ stands for the direct product and $\varepsilon$ is an
eight-component spinor whose components are left unspecified. Clearly,
condition (\ref{8spin}) reduces further the sixteen supersymmetry charges
to eight. Separating the eight components of $\varepsilon=(\varepsilon_7,
\varepsilon_1)$ where $\varepsilon_{7(1)}$ is a seven-(one-) dimensional
vector, we find that eq.(\ref{16SD}) reduces to the matrix equation
\bea
 {\cal O}\;\varepsilon\equiv\left(\begin{array}{cc}
\Psi_{imj}\dot{X}_m+\frac{\imath}2{{\cal X}^{mn}}_{ij}(-4)\{X_m,X_n\}
 & \dot{X}_i+\frac{\imath}2\Psi_{imn}\{X_m,X_n\}\\
-\left(\dot{X}_i+\frac{\imath}2\Psi_{imn}\{X_m,X_n\}\right)&0
\end{array} \right)\left(\begin{array}{c}
\varepsilon_7\\
\varepsilon_1\end{array} \right) = 0\label{delthe8}
\eea
The rather interesting fact here is that the matrix elements
${\cal O}_{8j}$ and ${\cal O}_{j8}$, $(j=1,..., 7)$ multiplying 
the $\varepsilon_1$-component are  the self-duality equations 
(\ref{sde}) in eight dimensions when the Euclidean time-parameter 
$t$ is replaced with $\imath t$ (Minkowski).
Thus, $\varepsilon_1$-component remains unspecified and there is
always one supersymmetry  unbroken for any eight-dimensional
solution of the self-duality equations. 

Let us now turn our discussion to the upper $7\times 7$ part
of the matrix equation (\ref{delthe8}). In general, the quantity
specifying these elements, namely
\beq
\Psi_{imj}\dot{X}_m+\frac{\imath}2{{\cal X}^{mn}}_{ij}(-4)\{X_m,X_n\}
\label{7x7}
\eeq
is {\it not} automatically zero. However, there is a particular case
--which turns out to be the most interesting one-- where the above
quantity is the self-duality equation itself. In fact, if we consider
only three-dimensional solutions of the equations, the `curvature'
factor ${\phi^{ij}}_{kl}$ is automatically zero while the tensor
${{\cal X}^{ij}}_{kl}$ simply becomes
\bea
{{\cal X}^{ij}}_{kl}=
{\Delta^{ij}}_{kl}=\frac 12(\delta_k^i\delta_l^j-\delta_l^i\delta_k^j)
&
{\rm for}\;\; {\phi^{ij}}_{kl}=0.
\eea
In this case, it can be easily seen that (\ref{7x7}) reduces to the
self-duality equations in three-dimensions. In this latter case, all 
eight supersymmetries survive.

We summarize this note discussing also the importance of the
supersymmetric self-duality configurations in three and seven
dimensions. The absence of a natural perturbative expansion
for the 11-d fundamental supermembrane prohibits so far the 
derivation of its low energy effective Lagrangian which is 
expected to contain  11-d, N=1 supergravity interacting 
with solitonic two- and five-branes in a duality symmetric way.
The Euclidean self-dual membrane configurations in three and
seven dimensions, after light-cone gauge fixing, provide 
non-perturbative minima of the action, which could survive
perturbative corrections if enough supersymmetries are left
intact. The quantum mechanical amplitudes calculated in 
supermembrane theory could be then determined by transforming
the path-integral integration around these minima into the
infinite moduli-space integration of the self-dual configurations 
of supermembranes. The best candidate for these seem to be
the three-dimensional integrable self-dual sector where
eight supersymmetries survive. The problem then is reduced
to find the moduli space and its integration measure of
the minimum action 3-d configurations. We hope to come
back to this problem in a future work.

% %  REFERENCES  


\newpage
\begin{thebibliography}{99}
% % 0
\bibitem{FL0}E.G. Floratos and G.K. Leontaris,
{\it Phys. Lett.} {\bf B 223} (1989) 153.
% % 1
\bibitem{1}T. Curtright, D. Fairlie and C. Zachos,
{\it  Phys. Lett.} {\bf B 405} (1997) 37.
% % 2
\bibitem{2} E.G. Floratos and G.K. Leontaris, {\tt hep-th/9710064,}
(to appear in {\it Nucl. Phys.} {\bf B 512} (1998) 445).
% % 3 
\bibitem{3}D. Fairlie, {\tt hep-th/9707190}; D. Fairlie and
T. Ueno, {\tt hep-th/9710079}; T. Ueno, {\tt hep-th/9801079}.
% % 4 
\bibitem{4}E.G. Floratos, G.K. Leontaris, A. Polychronakos and
R. Tzani, {\tt hep-th/9711044}, (to appear in {\it Phys. Lett.} {\bf B}).
% % 5
\bibitem{BST}E. Bergshoeff, E. Sezgin and P. Townsend, {\it Phys. Lett.}
{\bf B 176} (1987) 69; {\it Ann. Phys.} {\bf 185} (1988) 330;
E. Bergshoeff, E. Sezgin and Y. Tanii, {\it Nucl. Phys.}
{\bf B 298} (1988) 187.
% % 6 
\bibitem{DS}M. Duff and K.S. Stelle, {\it Phys. Lett.} {\bf B 253} (1991) 113.
% % 7
\bibitem{5aa}T. Kugo and P. Townsend,
 {\it Nucl. Phys.} {\bf B 221} (1983) 357;\\
S. Fubini and H. Nicolai, {\it Phys. Lett.} {\bf B 155} (1984) 431;\\
J. Evans, {\it Nucl. Phys.} {\bf B 298} (1988) 92.                           
% % 8
\bibitem{5a}B. de Wit and H. Nicolai, {\it Nucl. Phys.} 
 {\bf B 255} (1985)29.
% % 8a
\bibitem{ivan}B. de Wit and H. Nicolai, {\it Nucl. Phys.} {\bf B 231}
(1984)506;\\
T.A. Ivanova, {\it Phys. Lett.} {\bf B 315} (1993) 277.
% % 9
\bibitem{6}J. Harvey and A. Strominger, {\it Phys. Rev. Lett.} {\bf 66} 
(1991) 549;\\
T.A. Ivanova and A. Popov, {\it Lett. Math. Phys.} {\bf 24}
(1992)85;\\
M. G\"unaydin and H. Nicolai, {\it Phys. Lett.} {\bf B 351} (1995) 169.
% % 10
\bibitem{7}
M. Duff et al, {\it Phys. Lett.} {\bf B 412}
(1997) 281, {\tt hep-th/9706124}.
% % 11
\bibitem{7a}L. Baulieu, H. Kanno and I. Singer, {\tt hep-th/9705127}.
% % 12
\bibitem{wrap}B. de Wit, U. Marquard and H. Nicolai,
{\it Com. Math. Phys.} {\bf 128} (1990) 39;\\
K. Ezawa, Y. Matsuo and K. Murakami, {\tt hep-th/9705005};\\
B. de Wit, K. Peeters and J.C. Plefka, {\it Phys. Lett.}
{\bf B 409}(1997)
117;% {\tt hep-th/9710215}.
% % 13
\bibitem{NdW}B. de Wit, J. Hoppe and H. Nicolai, {\it Nucl. Phys.}
{\bf B 305 } (1988) 545.
% % 14
\bibitem{W} R.S. Ward, {\it Phys. Lett.} {\bf B234} (1990) 81.
% % 15
\bibitem{Gursey}
R. D\"undarer, F. G\"ursey and  Chia-Hsiung Tze, 
{\it J. Math. Phys.} {\bf 25} (1984) 1496.
\end{thebibliography}
\end{document}